\documentclass[usenatbib]{mn2e}
\usepackage{graphicx}
\usepackage{amsmath}
\usepackage{amssymb}
\usepackage{times}
\usepackage{mathtools}

\usepackage{morefloats}
\usepackage{etex}
\title[GAMA: Mergers and Properties]{Galaxy and Mass Assembly (GAMA): merging galaxies and their properties} 
\author[R. De Propris et al.]
{Roberto De Propris$^{1}$\thanks{E-mail: rodepr@utu.fi}, Ivan K. Baldry$^2$, Joss Bland-Hawthorn$^3$, Sarah Brough$^4$, 
\newauthor
Simon P. Driver$^{5,6}$, Andrew M. Hopkins$^4$,  Lee Kelvin$^7$, Jon Loveday$^8$, 
\newauthor
Steve Phillipps$^9$, Aaron S. G. Robotham$^5$ \\
$^1$ Finnish Centre for Astronomy with ESO (FINCA), University of Turku, V{\"a}is{\"a}l{\"a}ntie 20,
FI-21500 Piikki{\"o}, Finland \\ 
$^2$ Astrophysics Research Institute, Liverpool John Moores University, IC2, Liverpool Science Park, 146 Brownlow Hill, Liverpool, L3 5RF,UK\\
$^3$ Sydney Institute for Astronomy, School of Physics A28, University of Sydney, NSW 2006, Australia\\
$^4$ Australian Astronomical Observatory, PO Box 915, North Ryde, NSW 1670, Australia\\
$^5$ International Centre for Radio Astronomy Research (ICRAR), The University of Western Australia, 
35 Stirling Highway, Crawley, WA 6009, Australia\\
$^6$ Scottish Universities Physics Alliance (SUPA), School of Physics and Astronomy, University of St Andrews, North Haugh, KY169SS, UK\\
$^7$ Institut f{\"u}r Astro- und Teilchenphysik, Universit{\"a}t Innsbruck, Technikerstra{\ss}e 25, 6020 Innsbruck, Austria\\
$^8$ Astronomy Centre, University of Sussex, Falmer, Brighton BN1 9QH, United Kingdom\\
$^9$ H. H. Wills Physics Laboratory, University of Bristol, Royal Fort, Tyndall Avenue, BS8 1TL, United Kingdom}

\newcommand{\aj}{AJ}
\newcommand{\apj}{ApJ}
\newcommand{\aap}{A\&A}

\newcommand{\pasp}{PASP}
\newcommand{\apjs}{ApJS}

\newcommand{\mnras}{MNRAS}
\newcommand{\nat}{Nature}

\begin{document}

\date{Accepted. Received }

\maketitle

\label{firstpage}

\begin{abstract}

We derive the close pair fractions and volume merger rates for galaxies in the GAMA survey with 
$-23 < M_r  < -17$ ($\Omega_M=0.27$, $\Omega_{\Lambda}=0.73$, $H_0=100$ km s$^{-1}$ Mpc$^{-1}$) at $0.01 < z < 0.22$ (lookback time of $<2$ Gyr). The merger fraction is approximately 1.5\% per Gyr at all luminosities (assuming 50\% of  pairs merge) and the volume merger rate is $\approx 3.5 \times 10^{-4}$ Mpc$^{-3}$ Gyr$^{-1}$. We examine how the merger rate
varies by luminosity and morphology. Dry mergers (between red/spheroidal galaxies) are found to be uncommon and to decrease with decreasing luminosity. Fainter mergers are ‘wet’, between blue/disky galaxies. Damp mergers (one of each type) follow the average of dry and wet mergers. In the brighter luminosity bin ($-23 < M_r < -20$) the merger rate evolution is flat, irrespective of colour or morphology, out to $z \sim 0.2$. The makeup of the merging population does not appear to change over this redshift range. Galaxy growth by major mergers appears comparatively unimportant and dry mergers are unlikely to be significant in the buildup of the red sequence over the past 2 Gyr. We compare the colour, morphology, environmental density and degree of activity (BPT class) of galaxies in pairs to those of more isolated objects in the same volume. Galaxies in close pairs tend to be both redder and slightly more spheroid-dominated than the comparison sample. We suggest that this may be due to `harassment' in multiple previous passes prior to the current close interaction. Galaxy pairs do not appear to prefer significantly denser environments. There is no evidence of an enhancement in the AGN fraction in pairs, compared to other galaxies in the same volume
\end{abstract}

\begin{keywords}

galaxies: interactions --- galaxies: formation

\end{keywords}

\section{Introduction}

Mergers and interactions are believed to be among the primary pathways for galaxy formation and evolution: they are expected to drive 
star formation histories, morphologies, internal kinematics, chemical evolution and nuclear activity, among others, and reflect a long history 
of processing through different environments \footnote{The literature on this subject is very large and growing by the day; we cannot hope 
to be just or comprehensive in a brief introduction. However, \cite{Baugh2006} and the recent textbook by \cite{Mo2010} provide good entry 
points to this subject.}. The merger rate of galaxies, its evolution and its dependence on mass, luminosity, colour, morphology and environment 
(to name some) provide important clues to theories of galaxy formation and are an essential ingredient in simulations (e.g., \citealt{Murali2002,
Maller2006,Stewart2009,Perret2013} and references therein). 

However, it is actually difficult to estimate the galaxy merger rate and  compare with theoretical estimates (e.g.,\citealt{Berrier2006,Genel2009,
Williams2011,Moreno2013}). Several approaches have been applied to astronomical data and these often yield inconsistent results. Tidal 
features and sheets were early recognized by \cite{Zwicky1953, Zwicky1956} as signposts of galaxy interactions, following the pioneering 
opto-mechanical experiments by \cite{Holmberg1941}. Seminal work by \cite{Toomre1972} and \cite{Toomre1977} later established a 
connection between tidal disturbances and mergers, leading to a semi-empirical model for the formation of giant ellipticals from the merger 
of disks in the context of early Cold Dark Matter models (e.g., \citealt{Blumenthal1984}). More recent applications of this technique rely on 
objective measures of morphological anomaly in galaxy images, rather than visual classifications, such as the Gini coefficient
\citep{Abraham2003}, $M_{20}$ index \citep{Lotz2004} and the CAS method of \cite{Conselice2003} and Fourier-mode asymmetry
measurements \citep{Peng2012}. A companion paper by \cite{Casteels2013} also gives details on the application of CAS-like approaches 
to the GAMA survey. 

If a galaxy is to merge, it needs a close companion to merge with. The fraction of galaxies in close pairs therefore yields a proxy for the galaxy 
merger rate. With the availability of large redshift databases from giant redshift surveys it is now possible to identify close pairs in 3D space 
(position and relative velocity), such that, if bound, these galaxies will merge within the dynamical friction timescale ($\approx 1$ Gyr). The 
formalism and its application to the SSRS2 and CFRS surveys have been presented in \cite{Patton2000,Patton2002}; we have used this method 
to measure the local pair fraction and merger rate from the Millennium Galaxy Catalogue (MGC; \citealt{Liske2003,Driver2005}) in  
\cite{DePropris2005,DePropris2007}, where we also compared the results to asymmetry-based estimates \citep{DePropris2007}. Several 
other studies have searched for dynamically close pairs out to $z \sim 1$ (e.g. \citealt{Xu2004,Ryan2007,Lin2008,deRavel2009,
Bridge2010,Lopez2011} {\it et seq.}). 

Both methods have their own strengths and weaknesses. Morphology-based approaches require high resolution, high-quality imaging
and careful attention needs to be paid to contamination (e.g., edge-on galaxies, segmentation -- \citealt{DePropris2007,Jogee2009,
Casteels2013}). Dynamically close pairs are observationally expensive as highly complete redshift samples need to be acquired and 
they are still affected by projection effects \citep{Moreno2013}. Both approaches are also affected by uncertain timescales for the close
pairs to merge or the visibility of merger remnants above a given level of asymmetry or disturbance (e.g., \citealt{Lotz2011}). While the 
two methods yield results in reasonable agreement \citep{DePropris2007,Cotini2013}, with each other, asymmetries and other 
morphology-based methods generally identify objects in the process of merging or recent merger remnants, while close pairs  yield 
a measure of the 'progenitor' rate and may be more easily compared to theoretical predictions by identifying close pairs of dark
matter halos in simulations \citep{Genel2009,Lotz2011} as long as one understands the assumptions about the correspondence 
between dark matter halos and visible galaxies. 

Here we present a search for close pairs in the GAMA survey \citep{Driver2011}. This dataset is ideal for this measurement, as the survey 
has very high completeness ($>97\%$) and no bias against close companions (for instance, the fibre placement algorithm in the SDSS leads to
incompleteness at separations of $<55''$ and in the 2dF survey the close separation limit is about $30''$). We measure the pair fraction and 
volume merger rate as a function of absolute luminosity. We also split our samples into dry (between quiescent/spheroid-dominated galaxies), 
wet (star forming or disk-dominated objects) and damp (mixed) mergers and we consider their dependence on luminosity; we also measure 
the cosmic variance of the pair fraction and merger rate. We study the properties of galaxies in pairs and compare these (colours, morphologies, 
environment and activity) with those of more isolated objects (i.e.,not in our close pairs) in the same volume and with the same luminosity, to
understand how interactions have affected galaxies. The outline of this paper is as follows: we discuss the dataset and the methodology in 
section 2. We then present the pair fractions in section 3. In section 4 we consider the properties of galaxies in pairs vs. their parent samples. 
Finally, in section 5, we discuss the results in the context of theories of galaxy formation. Throughout this paper we assume the standard 
$\Lambda$CDM cosmology with $\Omega_M=0.27$, $\Omega_{\Lambda}=0.73$ and $H_0=100$ km s$^{-1}$ Mpc$^{-3}$.

\section{Dataset and Methodology}

The dataset we use is the Galaxy and Mass Assembly (GAMA) redshift survey. This is fully described in Driver et al. (2011, see
\citealt{Baldry2010} for the input catalogue and \citealt{Hopkins2013} for details of the spectroscopic analysis); in the following, 
we give a brief summary of its main properties as used in this paper. GAMA as used in this paper is equivalent to the GAMA-I
sample obtained during the first 3 years of the survey and frozen for internal team use and consists of three 48 deg$^2$ 
regions on the celestial equator, at RA of 9h, 12h and 14.5h, containing photometry over a wide range of wavelengths, including 
vacuum UV from GALEX, $u,g,r,i,z$ from the SDSS \citep{York2000,Ahn2012}, $(Y)JHK$ from 2MASS and UKIDSS, mid-IR 
from the WISE survey at 3.6, 4.5, 12 and 22 $\mu$m, as well as other sources of archival photometry (e.g., HERSCHEL, Planck). 
Spectroscopy to a limiting magnitude of $r=19.4$ ($r=19.8$ for the 12h region) has been obtained for over 120,000 galaxies with 
the AAOmega multi-fibre spectrograph on the Anglo-Australian Telescope, supplementing existing datasets from the 2dF Galaxy 
Redshift Survey \citep{Colless2001}, the SDSS and others, to reach a high degree ($> 97\%$) of spectroscopic completeness 
\citep{Hopkins2013}. Each region in the sky is surveyed multiple times, which aids in reaching high completeness at small angular 
separations. Because of this, GAMA is ideal to carry out a close pair analysis of galaxy populations to measure the merger rate and 
explore the influence of interactions and mergers on galaxy properties. 

We search for pairs in GAMA using the methods developed by \cite{Patton2000,Patton2002} and used in our previous work on the subject
\citep{DePropris2005,DePropris2007,DePropris2010}. Following previous studies, we define galaxies as being in a close pair if their 
projected separation on the sky is $5 < r_ p < 20$ kpc (the lower limit is imposed to avoid selecting high surface brightness regions within 
galaxies, such as HII regions) and if their velocity difference is $\Delta V < 500$ km s$^{-1}$. If these objects are truly bound, we expect 
that they will merge, by dynamical friction in $< 1$ Gyr (the actual timescales are likely to be somewhat longer, see below for details).

\begin{figure}
\includegraphics[width=0.5\textwidth]{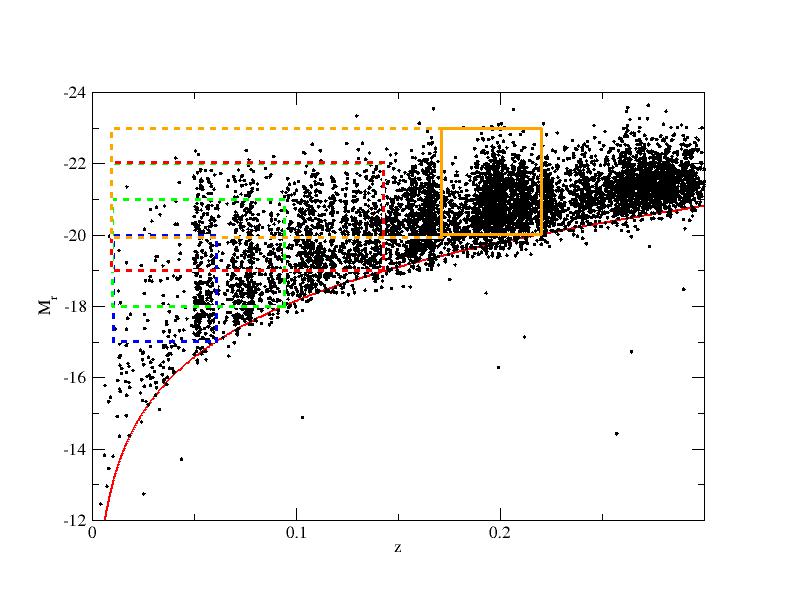}
\caption{Galaxies in the GAMA survey (we plot only a small subset for reasons of clarity) 
plotted as absolute magnitude vs. redshift (top panel). The coloured boxes on the plot show the volume limited 
samples that we search for pairs in. The limits in magnitude and redshifts are shown in Table 1. The 
two orange boxes show the adopted samples for galaxies with $-23 < M_r < -20$ 
and $0.0100 < z < 0.1724$  and $0.1724 < z < 0.2193$, respectively. The two redshift intervals are 
chosen to yield approximately equal volumes. The red box is for galaxies with $-22 < M_r  < 
-19$ at $0.0100 < z < 0.1442$; the green box is for galaxies with  $-21 < M_r < -18$ at 
$0.0100 < z < 0.0939$ and the blue box is for galaxies with $-20 < M_r  < -17$ at $0.0100 < z < 0.0607$. 
The thick red line shows the predicted absolute magnitude for a slowly evolving galaxy (discussed in text) 
having $r=19.4$ at $z=0$. }
\label{fig:fig1}
\end{figure}

Fig. 1 shows the absolute magnitude of galaxies in GAMA (only a subset is shown for clarity) as a function of redshift. Here, all magnitudes 
and $k+e$ corrections are as given in the GAMA-I catalogue \citep{Driver2011}. The first step in our analysis is to select 
galaxies within a series of volume-limited boxes in absolute magnitude and redshift space. In order to measure a pair fraction, we need to 
restrict ourselves to a relatively small range of luminosities, as we need objects to have similar clustering properties \citep{
Patton2000,Patton2008} and the clustering length is known to vary as a function of galaxy luminosity \citep{Norberg2001}. However, we 
also need our samples to contain enough galaxies to determine the pair fraction to sufficient accuracy, given that only $<3\%$ of galaxies 
(based on previous work) are likely to fulfil our definition of a close pair. As in \cite{Patton2000,Patton2002,DePropris2005,DePropris2007}, 
we draw 'volume-limited' samples spanning 3 mag. in luminosity (Fig.~1). Potentially, this includes mergers with luminosity ratios of up to
1:10, but the majority of close pairs will have luminosity ratios of $<1:4$ and are therefore `major' mergers (see below). Table~1 shows the 
luminosity and redshift ranges we study in this work, as well as the numbers of objects in each volume-limited box and the number of galaxies
in close pairs (this need not be an even number, because of possible triplets). Note that for the brighter luminosity bin we consider, we are 
able to draw two samples, at $<z> = 0.09$ and $<z>=0.19$, and study the evolution of the merger rate between these two redshifts for 
galaxies with $-23 < M_r < -20$.

Because our sample (like all samples) is luminosity-limited we need to impose a lower absolute magnitude limit to our search, rather than
using all galaxies in the appropriate luminosity range. This corresponds to the luminosity of a slowly evolving galaxy having the apparent 
magnitude limit of the survey ($r=19.4$) at all redshifts considered (thick red line in Fig.~1). If this is not done, it is possible for galaxies to 
enter or leave the sample depending on their star formation history \citep{Patton2000}. We use a \citealt{Bruzual2003} model with $z_f=3$, 
$Z=Z_{\odot}$ and $\tau=1$ Gyr, evolving passively, as a template for the slowest evolving object we are likely to encounter. When drawing 
the volume-limited boxes shown in Fig.~1, this model sets the high redshift limit for each luminosity interval we examine. 

Each pair consists of a primary and secondary member; in this case we choose to make the two samples coincide so each galaxy is counted 
twice, once as a central and next as a satellite of the other. Following \cite{Patton2000, Patton2002} and our previous work, the number of 
close companions per galaxy is:

$$N_c={{\sum_i^{N_1} w^i_{N_1} N_{c_i}} \over {\sum_i^{N_1} w^i_{N_1}}} , $$

where the sums are over the $i=1,...,N_1$ galaxies in the primary sample, and the total companion luminosity is:

$$L_c={{\sum_i^{N_1} w^i_{L_1} L_{c_i}} \over {\sum_i^{N_1} w^i_{L_1}}} ,$$

Here $N_{c_i}$ and $L_{c_i}$ are the number and luminosity (respectively) of galaxies in the secondary sample which are close companions
(by the above definition) to the $i^{th}$ galaxy in the primary sample:

$$N_{c_i}=\sum_j w^j_{N_2} = \sum_j {w^j_{b_2} w^j_{v_2} \over S_N(z_j)}, $$

and

$$L_{c_i}=\sum_j w^j_{N_2}L_j = \sum_j {w^j_{b_2} w^j_{v_2} \over S_L(z_j)} L_j ,$$

where the sums run over the $j$ secondary galaxies that fulfil the criteria of being dynamically close to the $i^{th}$ primary galaxy. 
The raw pair fractions are then weighted to correct for sources of spatial and spectroscopic incompleteness (e.g., areas where no
fibers were placed such as within the halos of bright stars, or limits to slit placement for close companions).

The components $S_N(z)$ and $S_L(z)$ are weights to correct for the change in density of the secondary galaxies as a function of
redshift due to the apparent flux limit of the sample and convert a luminosity limited sample into what would be expected from a
volume-limited sample. They are calculated as the ratio of the integrated luminosity function over the  luminosity sampled to the integrated 
luminosity function in a hypothetical volume-limited sample over the same redshift. In our case, as we are using volume-limited 
subsets of the data (see Figure 1), these components are equal to 1 (no weight). 

The other two weights, $w_{b_2}$ and $w_{v_2}$ correct for boundary effects. The first one takes care of areas (within the search radius) 
around each galaxy where no companion can be found because they lie beyond the edges of the survey on the sky or within areas occulted 
by bright stars, `drill holes' (for guide probes), image defects and satellite trails or other cosmetic issues. This weight is the reciprocal of
the fraction of the $\pi r_p^2$ area around each galaxy which is occulted in some fashion. The second weight is for galaxies close (within 
500 km s$^{-1}$) to the redshift limits of each volume-limited box, and accounts for possible companions missed because they lie just beyond 
these boundaries. As in \cite{Patton2000,Patton2002} the value of this weight is set at 2. Similar weights need to be applied to the primary
sample as well, to account for missed primaries in the same manner (recall that both primary and secondary samples are the same):

$$ w^i_{N_1}=w^i_{b_1} w^i_{v_1} S_{N_1}(z),$$

$$ w^i_{L_1}=w^i_{b_1} w^i_{v_1} S_{L_1}(z),$$

these weights being the reciprocals of those applied to the secondary sample $w^j_{b_2}$, $w^j_{v_2}$ \citep{Patton2000}.

Although GAMA is highly complete and is intended to have no pair bias at close angular separations, we need to correct for potential pairs 
missed because of redshift incompleteness, especially at small separations. We carry this out in the following manner: we first select potential 
pairs with galaxies that have no redshifts by the projected separation criterion only, and if such a pair exists, we assign to the galaxy the same 
redshift and $k+e$ corrections as its primary and require that the secondary galaxy falls within the selection criteria in luminosity. We then 
estimate the fraction of 'true' pairs in this sample by carrying out the same analysis on the photometric sample (only) and on the redshift 
sample (only) separately, by using only the projected separation criterion. Since we know the true pair fraction in the redshift sample, we 
can use the ratio between the pair fractions in the 'photometric' and 'redshift' samples to derive an incompleteness correction. This is similar
to the approach used by \cite{DePropris2005} for the MGC and by \cite{Masjedi2006,Masjedi2008}.  Naturally, this assumes that pairs are 
missed in a random fashion and potential companions have the same redshift distributions as the overall GAMA survey. This completeness 
correction amounts to $\sim 1\%$ to $4\%$ between $M_r=-23$ and $M_r=-17$, respectively, with some increase towards lower luminosities, 
in agreement with GAMA's high overall completeness and broad lack of bias at small separations.

An alternative approach is to select galaxies by stellar mass rather than luminosity. For the GAMA dataset the galaxy stellar mass was 
calculated by \cite{Taylor2011} who gives a formula involving their absolute  $r$-band luminosity and rest-frame $g-i$ colour. However,
this is found to reject a significant fraction of the sample, as the stellar mass determinations are model-dependent (cf. the equivalent
figure in Robotham et al. (2014, submitted to MNRAS). Our aim in this paper is not only to measure the pair  fraction and derived merger 
rate, but also to consider how these quantities depend on luminosity,morphology, colour and environment and on how the properties of 
galaxies in pairs compare to those of more isolated systems in the same volume. Therefore we will restrict ourself to the numerically larger 
luminosity-selected samples. A pair analysis of GAMA data by stellar mass can be found in the companion paper by Robotham et al. (2014), 
where some of the selection issues are addressed with reference to simulations and mock catalogues. Nevertheless, this latter analysis does 
not concern itself with the dependencies on luminosity, colour, morphology and environment which are among the topics explored here. 

\section{Pair Fractions}

Table 1 shows the main results of our analysis. In column order, the table contains: the luminosity ranges, redshift ranges, number of galaxies 
in each volume-limited box and number of close pairs, the completeness corrected pair fractions and the volume merger rates. In general,
these objects are major mergers. Around 2/3 of all objects, in all the samples considered, have luminosity ratios of 1:2 or greater, and about
20\% have luminosity ratios between 1:2 and 1:4. 

\begin{table*}
 \begin{minipage}{0.8\textwidth}
  \caption{Pair fractions and volume merger rates}
  \begin{tabular}{cccccc}
  \hline
   Luminosity Range & Redshift Range & N$_{obj}$ & N$_{pairs}$ &Pair Fraction (\%) & Volume Merger Rate \\
   mag & & & & & $10^{-4}$ Mpc$^{-3}$ Gyr$^{-1}$ \\
 \hline
$-23 < M_r  < -20$ & $0.0100 < z < 0.1724$  & 19816 & 644 & $3.30 \pm 0.10$  & $3.33 \pm 0.11$ \\
$-23 < M_r  < -20$ & $0.1724 < z < 0.2193$  & 17093 & 494 & $ 3.03  \pm 0.14$  & $2.60 \pm 0.12$ \\
$-22 < M_r  < -19$ & $0.0100 < z < 0.1442$ & 20936 & 699 & $ 3.39  \pm 0.13$  & $4.39 \pm 0.17$ \\
$-21 < M_r < -18$ & $0.0100 < z < 0.0939$ &  8084 & 256 & $ 3.24 \pm 0.21$  & $4.26 \pm 0.28$ \\
$-20 < M_r < -17$ & $0.0100 < z < 0.0607$ &  3151 & 56 & $ 1.84  \pm 0.24$ & $2.51 \pm 0.33$ \\
 \hline
\end{tabular}
\end{minipage}
\label{tab:tab1}
\end{table*}

The volume merger rates in this table are calculated as:

$$R_{mg}=N_c n(z) 0.5 p_{mg} T_{mg}^{-1},$$

Where $N_c$ is the close pair fraction (as in Table 1), $n(z)$ the volume density of galaxies (calculated from the completeness-corrected
number of galaxies in each sample box), the factor of 0.5 takes care of the fact that each galaxy in a pair is counted twice (once as a primary 
and the other as a secondary), $p_{mg}$ is the probability that the pair will truly merge (here 0.5, as in previous work and as supported by
observations \citealt{DePropris2007,Jian2012,Casteels2013}) and $T_{mg}$ is the merger timescale. This is calculated as follows: for each 
luminosity interval we compute the mean stellar mass of the pairs (using the colours and the expression from \citealt{Taylor2011}) and then 
adopt the fitting formulae of \cite{Kitzbichler2008} and \cite{Boylan2008} for the merger timescales. These range between 0.5 and 1.5 Gyr 
for galaxies in higher and lower luminosity bins, respectively. These values are in good agreement with those estimated by \cite{Boylan2008}, 
as well as those stated by \cite{Lotz2010} for the visibility of morphologically-selected merger remnants and the compilation in \cite{Lotz2011}. 
\cite{Patton2008} also use this approach, and derive a volume merger rate of $6 \times 10^{-5}$ $h^3$ Mpc$^{-3}$ Gyr$^{-1}$ per mag.
fpr galaxies with $-22 < M_r + 5 \log h < -18$ and 1:2 luminosity ratio, assuming a single timescale of 0.5 Gyr for all objects. If they adopted
the fitting formulae by \cite{Kitzbichler2008} these would result in merger timescales of 0.8 to 3.2 Gyr depending on galaxy luminosities, which
are  somewhat longer than ours. Compared to the more conventional  estimates from dynamical friction, the merger timescales are about 
30\% longer, but as pointed out by \cite{Jiang2008} these are due to the poor performance of the simple estimate of the Coulomb logarithm. 
As we give explicit values for the raw pair fraction (with completeness corrections), as well as details on the luminosities and volumes 
sampled, future studies or theoretical comparisons can adopt any appropriate timescale and apply it to the data presented here.

\cite{Jian2012} show that the merger timescale depends sensitively on the mass ratio of mergers, being shorter for equal masses and 
considerably longer for more minor mergers. Most of our pairs, on the other hand, have similar luminosity ratios (about 4/5 of all objects 
are $<1:4$ mergers). The timescales we calculate above are therefore appropriate for nearly equal mass mergers as inn \cite{Jian2012}, 
although our values may be a slight underestimate of the actual merger timescale. We tabulate the derived volume merger rates in units 
of $10^{-4}$ Mpc$^{-3}$ Gyr$^{-1}$ in Table~1. These amount to mean luminosity accretion rates of $0.5 \times 10^8 < L/L_{\odot} < 
5 \times 10^8$ per Gyr for galaxies with $-23.0 <  M_r  < -17$. If we use the stellar masses computed following \cite{Taylor2011}, these 
luminosity accretion rates correspond to stellar mass accretion rates of $O(10^9)$ $M_{\odot}$ Gyr$^{-1}$, equivalent to growth factors 
of $\approx 5\%$ of the current mass of each galaxy per Gyr.

\begin{figure}
\includegraphics[width=0.5\textwidth]{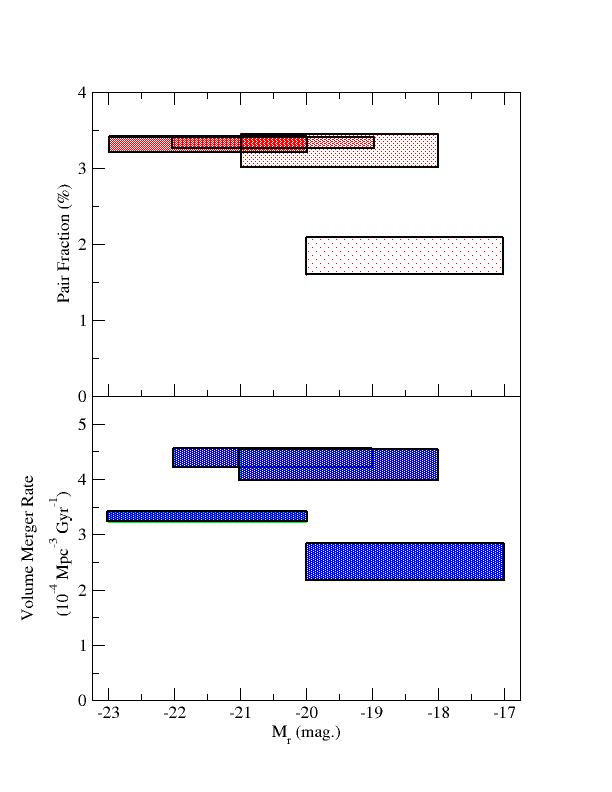}
\caption{Pair fractions (top panel, red boxes) and volume merger rates (bottom panel, blue boxes) as a function of luminosity. 
The width of the boxes represents the width of the respective luminosity bin, while the height of the boxes represents the 1$\sigma$ 
error as in Table~1 above.}
\label{fig:fig3}
\end{figure}

We present these quantities in Fig.~\ref{fig:fig3}. The pair fraction (top panel) is a nearly constant 3\% at all luminosities. Note that the 
actual {\it merger fraction} is 1/2 of these values, if we assume that only 50\% of close pairs actually merge. The bottom panel of this
figure shows the equivalent for volume merger rates. In both cases we observe a decline in the merger fraction (and rate) in the lower
luminosity bin, and a possible decrease at high luminosities as well; the former is in agreement with observations by \cite{Patton2008}.
These trends are more significant for volume merger rates, because of the different volume densities and merger timescales for each 
luminosity bin. We are also able to measure the pair fraction and volume merger rate for galaxies in the brightest luminosity bin within 
two redshift intervals (centred on $z=0.09$ and $z=0.19$) to measure the evolution of the merger rare since $z < 0.2$ (approximately 2 Gyr 
in the past). We do not show the higher redshift datapoints in ~\ref{fig3} and subsequent figures to avoid confusion. Our measurements are
consistent with a flat or even declining merger rate evolution, in agreement with previous observations for similarly massive galaxies
\citep{Lin2004,Kartaltepe2007,Lin2008,deRavel2009} at higher redshifts, which are also consistent with $R_{mg} (z) \sim (1+z)^{\sim 0}$
\citep{Lotz2011}.

\subsection{Dependence on colour and morphology}

We now consider how the merger rate depends on the colour of galaxies. For example, mergers between gas-less galaxies ('dry' mergers, 
with galaxies already on the red sequence) have been proposed as a possible mechanism to grow the red sequence in mass without increasing 
the intrinsic scatter in the colour-magnitude relation and other scaling factors \citep{Bell2004,vanDokkum2005,Faber2007,Skelton2012}. 
Obviously, we have no knowledge of the gas content of these galaxies, but we can use the $u-r$ colour to separate quiescent and actively 
star-forming galaxies, with the dividing line at $u-r=2.2$ (e.g., \citealt{Strateva2001,Baldry2004}). This is acceptable for our purposes, as 
there is a moderate degree of correlation between colour and gas content, in the sense of red galaxies being more gas-poor 
\citep{Toribio2011}. Here 'dry' mergers are between two 'red' galaxies, 'wet' mergers between two 'blue' galaxies and 'damp' mergers 
contain one member of each. 

Another possible definition uses morphology. Galaxy profiles (single S{\'ersic} models) have been derived for GAMA galaxies in \cite{Kelvin2011}.
We define galaxies with $n > 2.5$ as spheroid-dominated and those with $n < 2.5$ as disk-dominated. More spheroid-dominated galaxies are 
known to be more gas poor and we can therefore use this as a proxy of relative 'dryness'. Here, dry mergers are those between 
two spheroid-dominated galaxies, while wet mergers are those between two disk-dominated objects, with damp mergers having one member of 
each class. The derived pair fractions and volume merger rates for dry, wet and damp mergers (selected by colour and morphology, as
above) are tabulated in Tables 3 and 4 respectively.  

\begin{table*}
 \center
 \tabcolsep=0.1cm
 \begin{minipage}{\textwidth}
  \caption{Pair fractions for dry, wet and mixed mergers}
  \begin{tabular}{lccccccc}
  \hline
   Luminosity Range & Redshift Range & Blue Pairs & Red Pairs & Mixed Pairs & Disk Pairs & spheroid Pairs 
   & spheroid/Disk Pairs \\
 \hline
\small
$-23 < M_r < -20$ & $0.0100 < z < 0.1724$ &  $0.0101 \pm 0.0003$ &  $0.0132 \pm  0.0003$ &  $0.0097  \pm  0.0003$ & $0.0056 \pm 0.0002$  & $ 0.0156  \pm 0.0005$ &  $0.0118 \pm  0.0004$ \\
$-23 < M_r < -20$ & $0.1724 < z < 0.2193$ &  $0.0120  \pm 0.0005$ &  $0.0061 \pm 0.0003$ &  $0.0126 \pm  0.0006$  & $0.0061  \pm  0.0003$  & $0.0108 \pm 0.0005$ &  $0.0148 \pm  0.0007$ \\
$-22 < M_r < -19$ & $0.0100 < z < 0.1442$ & $0.0135  \pm 0.0005$ &  $0.0097 \pm  0.0004$  & $0.0107 \pm 0.0004$  & $ 0.0097 \pm  0.0004$ &  $0.0115 \pm  0.0004$ & $0.0128 \pm 0.0005$ \\
$-21 < M_r < -18$ & $0.0100 < z < 0.0939$ & $0.0177  \pm 0.0011$ & $0.0047 \pm  0.0003$ &  $0.0099 \pm 0.0006$  &  $0.0138 \pm 0.0009$ & $0.0058  \pm 0.0004$ & $0.0128   \pm 0.0008$ \\
$-20 < M_r < -17$ & $0.0100 < z < 0.0607$ & $ 0.0130 \pm  0.0017$ & $0.0013 \pm  0.0002$ & $ 0.0040 \pm 0.0005$ & $0.0138  \pm 0.0018$ &  $0.0007 \pm  0.0001$ &  $ 0.0039 \pm   0.0005$\\
  \hline
\end{tabular}
\end{minipage}
\label{tab:tab4}
\end{table*}

\begin{table*}
 \center
 \begin{minipage}{\textwidth}
  \caption{Volume merger rates (units of $10^{-4}$ Mpc$^{-3}$$h^{-3}$ Gyr$^{-1}$) for dry, wet and mixed mergers}
  \begin{tabular}{lccccccc}
  \hline
   Luminosity Range & Redshift Range & Blue Pairs & Red Pairs & Mixed Pairs & Disk Pairs & spheroid Pairs 
   & spheroid/Disk Pairs \\
 \hline
$-23 < M_r < -20$ & $0.0100 < z < 0.1724$ & $1.02 \pm 0.03$ & $1.32 \pm 0.03$ & $0.98 \pm 0.03$ & $ 0.56 \pm 0.02$  &  $1.56 \pm  0.05$ &  $1.18   \pm 0.04$ \\ 
$-23 < M_r < -20$ & $0.1724 < z < 0.2193$ & $1.03 \pm 0.04$ & $0.52 \pm 0.03$ & $1.08 \pm 0.05$ & $ 0.52 \pm 0.03 $ &  $ 0.85  \pm  0.04 $ & $   1.27   \pm  0.06  $ \\  
$-22 < M_r < -19$ & $0.0100 < z < 0.1442$ & $1.75 \pm 0.06$ & $1.26 \pm 0.05$ & $1.39 \pm 0.06$ & $1.26 \pm 0.05 $ & $1.49  \pm  0.05$ &  $1.66  \pm   0.06$ \\
$-21 < M_r < -18$ & $0.0100 < z < 0.0939$ & $2.37 \pm 0.14$ & $0.62 \pm 0.04$ & $1.30 \pm 0.08$ & $1.81 \pm 0.12 $ & $ 0.76 \pm  0.05 $ &  $1.68 \pm    0.11 $ \\
$-20 < M_r < -17$ & $0.0100 < z < 0.0607$ & $1.77 \pm 0.23$ & $0.18 \pm 0.03$ & $0.55 \pm 0.07$ &  $1.89 \pm 0.25 $ &  $ 0.10  \pm  0.01$ &  $ 0.53  \pm  0.07$ \\
\hline
\end{tabular}
\end{minipage}
\label{tab:tab5}
\end{table*}

\begin{figure*}
\includegraphics[width=\textwidth]{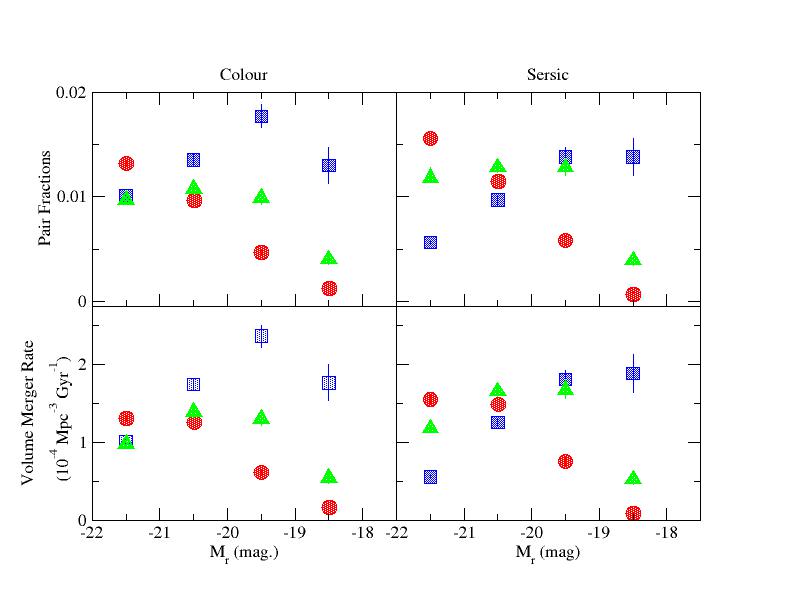} 
\caption{Pair fractions (top panels) and volume merger rates (bottom panels) for galaxies as a function of
luminosity. The points correspond the midpoint of each 3 mag. luminosity bin, as shown in Fig. 2 and Tables 1 and 2. 
We do not show the luminosity range of the bins here to avoid confusion. We separate pairs into 'dry mergers' (between 
two galaxies both with $u-r > 2.2$ on the left or having $n > 2.5$ in the right-hand panels) plotted as red circles, 'wet 
mergers'  ($u-r < 2.2$; $n < 2.5$), plotted as blue squares and 'damp' mergers (one galaxy in each colour range or S{\'ersic} 
index) as green triangles. Error bars are generally smaller than the symbols.}
\label{fig:fig4}
\end{figure*}

Figure~\ref{fig:fig4} shows the pair fractions and volume merger rates for galaxies, split into dry, wet and damp mergers. For the brighter 
luminosity range, we also tabulate separate results for the samples at $<z>=0.08$ and at $<z>=0.19$ (but note that as in the previous figure 
we do not show this datapoint in Fig.~\ref{fig:fig4} to avoid confusion). 

We observe dry mergers to be rare in our most luminous sample at $z < 0.2$, in agreement with previous results by \cite{Masjedi2006,
Masjedi2008,Wen2009,Robaina2010} and \cite{Jiang2012}. The dry merger rate also decreases monotonically as a function of luminosity. 
There are very few faint dry mergers. It appears therefore unlikely that the red sequence can be built by this process at least since $z \sim 0.2$ 
as probed in this study. Most mergers at the faint end are wet and their relative contribution increases with decreasing $M_r$. Damp mergers 
are intermediate between the behaviour of red and blue pairs. This is true for samples selected by colour or S{\'ersic} index. The intermediate 
behaviour of damp mergers suggests that galaxies in pairs are selected almost at random from the parent population and therefore that 
presence in a close pair does not strongly affect the  properties of the galaxies. We will examine this in greater detail below. In the brightest 
luminosity bin we also find evidence of only flat evolution in the merger rate, irrespective of colour or morphology, since $z \sim 0.2$, suggesting 
that there has been no change in the makeup of merging pairs (at least for massive galaxies) over the last $\sim 2$ Gyr of the history of the 
Universe. 

\subsection{Cosmic Variance}

\begin{figure*}
\includegraphics[width=\textwidth]{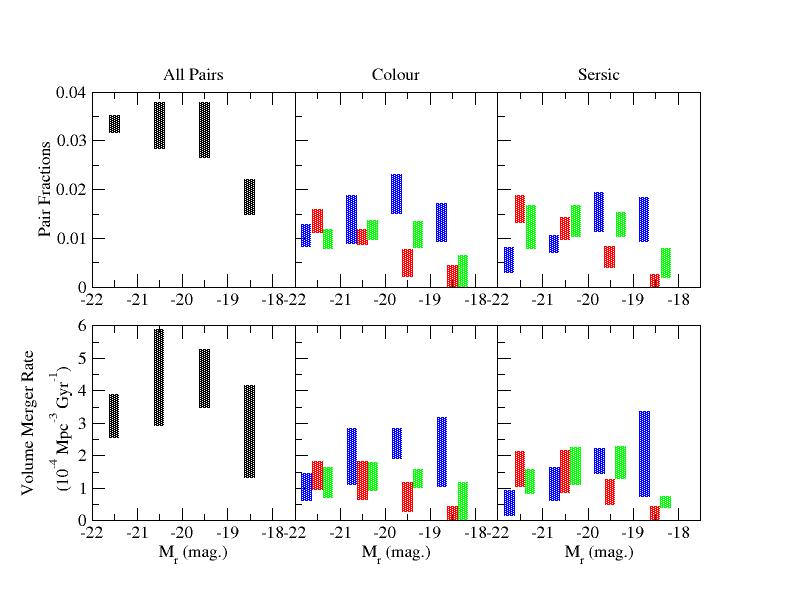}\\
\caption{Pair fractions (top panels) and volume merger rates (bottom panels) for galaxies in each individual GAMA region. The left-hand
panels show all objects, the middle panels plot pairs selected by colour and the right-hand panels pairs selected by their S{\'ersic}
index. The size of the bars shows the range of values for the pair fraction (without the error bars) and volume merger rate across 
all regions and hence the cosmic variance in these quantities. The black bars are for all objects, while red, blue and green bars are for
dry, wet and damp mergers, respectively, the same colour scheme as adopted in previous and subsequent figures. We have applied
small shifts in luminosity (+0.25 mag. for the blue bars and --0.25 mag. for the green bars) about the midpoint of each bin (as in
Tables 1 and 2) so that the bars do not overlap.}
\label{fig:fig6}
\end{figure*}

As GAMA consists of three separate regions, we calculate the pair fractions and volume merger rates for galaxies in each separate region to 
estimate the cosmic variance. This is shown in Fig.~\ref{fig:fig6} for the full sample as well as for dry, wet and damp mergers (by colour 
and morphology). We have applied small shifts in luminosity in this figure for reasons of clarity. The cosmic variance over volumes of 
$\sim 10^5$ $h^{-3}$ Mpc$^3$ is approximately a factor of 2, in reasonable agreement with the estimate by \cite{Lopez2014} from the 
ALHAMBRA survey. 

\section{The Properties of Galaxies in Pairs}

We now compare the properties of galaxies in pairs (colour, morphology, environmental density and degree of activity, such as star formation 
or AGNs, measured from the BPT index of \citealt{Baldwin1981}) to those of more isolated (i.e., not in close pairs) galaxies in the 
volume-limited region from which the pairs are drawn. We divide all galaxies into bins of colour, morphology, environmental density
and BPT class and compute the fractions of galaxies in each bin, for objects in close pairs and for their parent sample (more isolated
systems) in the same volume. We then calculate the fractional difference (pairs -- parent sample) in each colour (morphology, 
environment, BPT class) bin. A negative value means that there is a relative deficit of galaxies in close pairs in the appropriate colour 
(morphology, etc.) bin (i.e., they are less frequent) and vice-versa for a positive value. In this way we can asses how membership in a close
pair and the on-going interaction affects the properties of galaxies (relative to similar objects in the same volume). We show the results
of this analysis in Fig.~5--8 below.

\begin{figure*}
\includegraphics[width=\textwidth]{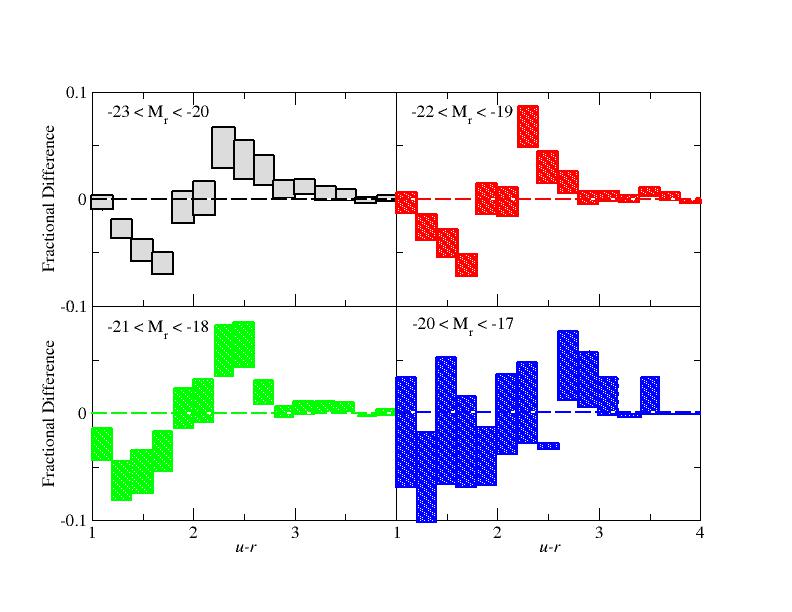}
\caption{Fractional difference in the distribution of $u-r$ colours for galaxies in pairs and the parent sample. A negative
number means that there are fewer pairs, in proportion, in the colour bin considered, with respect to the distribution of the
volume-limited box in Fig.~1 and vice versa. The length of each bar shows the $1\sigma$ range of the fractional
difference at each colour, while the width of the bar shows the colour range we study. The luminosity ranges considered
are found in the figure legend.}
\label{fig:fig8}
\end{figure*}

The colour distribution of galaxies is bimodal, as is well-known \citep{Strateva2001,Baldry2004}, with red galaxies becoming
more prominent in the higher luminosity bins. In all cases, except the faintest luminosity range where most galaxies are blue,
we note that pairs are overrepresented in the red peak and less frequent in the blue peak (Fig.~\ref{fig:fig8}). This argues that 
presence in a close pair does not trigger but rather tends to quench star formation. Similar claims of a greater frequency of red 
galaxies in pairs were earlier reported by \cite{Darg2010a} and \cite{Ellison2010}. \cite{Chou2012} also find that close pairs tend
to quench, and not enhance, star formation in their sample. \cite{Bergvall2003} argued that there is only a modest enhancement
in star formation rates for a sample of highly interacting objects. In our previous study of pairs in the MGC \citep{DePropris2005} we 
noticed a similar excess of red pairs, as well as an excess of blue pairs (and a deficit of green valley systems in pairs), which is not
seen here. \cite{Darg2010a} suggest that the apparent increase in the fraction of blue galaxies in pairs in some previous studies
may be due to greater visibility of gas-rich, star-forming objects, especially if pairs and mergers are selected by morphology.

The distribution of S{\'ersic} indices shows a deficit of 'disky' galaxies among pairs, and a slight excess of more spheroid-dominated
objects.  This may imply a degree of morphological evolution as part of the merger process. We found a similar trend in MGC 
data \citep{DePropris2005} with an excess of E/S0 and Sc/Sd galaxies in pairs compared to the MGC sample, and a deficit of 
intermediate spirals (see also \citealt{Darg2010a}). We also concluded (as do \citealt{Darg2010a}) that this is due to a combination
of morphological evolution and induced star formation. Most of these changes (in colour and morphology) appear to take place 
close to the 'transition' colour of $u-r=2.2$ or S{\'ersic} index of $n=2.5$. This may indicate that mergers and interactions do not
cause abrupt changes in star formation rates or morphologies, but might simply accelerate or enhance an on-going process of
secular evolution (as in the 'harassment' scenario by \citealt{Moore1996}). 

However, we cannot exclude the possibility that environment plays a role. \cite{Patton2011} notes a similar trend of increasing spheroid
fractions among galaxies in pairs and attributes it to the denser environments in which pairs may reside, yielding larger spheroids because 
of the morphology-density relation. \cite{Deng2013} use a volume-limited sample in the SDSS and reach similar conclusions. We therefore 
look at the environmental density of galaxies in pairs vs. other (more isolated) galaxies in the same volume. Here, environmental density 
is defined using an index developed by \cite{Haas2012} who use the distance to the $7^{th}$ nearest neighbour within 1000 km s$^{-1}$ 
and which is believed to provide a halo-independent measure of the environment of galaxies; this is more appropriate to our purposes (and 
less model-dependent) as we are chiefly interested in exploring whether differences in colour or morphology can be attributed to environmental 
effects. We derive this index from our GAMA data by applying the recipe of \cite{Haas2012}: a shorter distance (i.e., closer 7$^{th}$ 
neighbour) implies a denser environment and vice-versa (higher value of this index implies a less dense environment). Generally, GAMA 
samples a variety of environments, but most galaxies lie in the field or groups, with only one large cluster known in the GAMA fields. 

\begin{figure*}
\includegraphics[width=\textwidth]{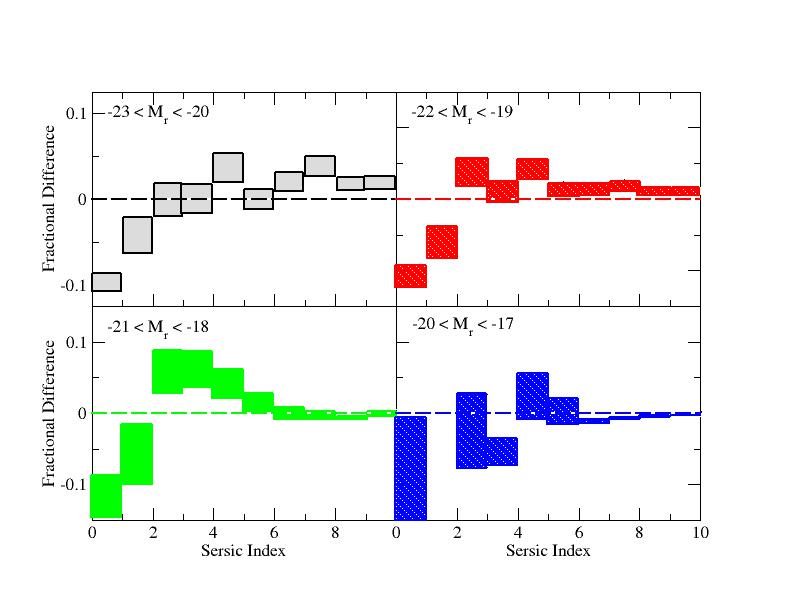}
\caption{Fractional difference in the distribution of S{\'ersic} indices for galaxies in pairs and the parent sample. A negative
number means that there are fewer pairs, in proportion, in the  bin considered, with respect to the reference distribution of
all galaxies in the volume-limited boxes of Fig.~1, and vice versa. The length of the bar shows the $1\sigma$ range
of the fractional difference while its width shows the range of S{\'ersic} indices we consider. }
\label{fig:fig9}
\end{figure*}

\begin{figure*}
\includegraphics[width=\textwidth]{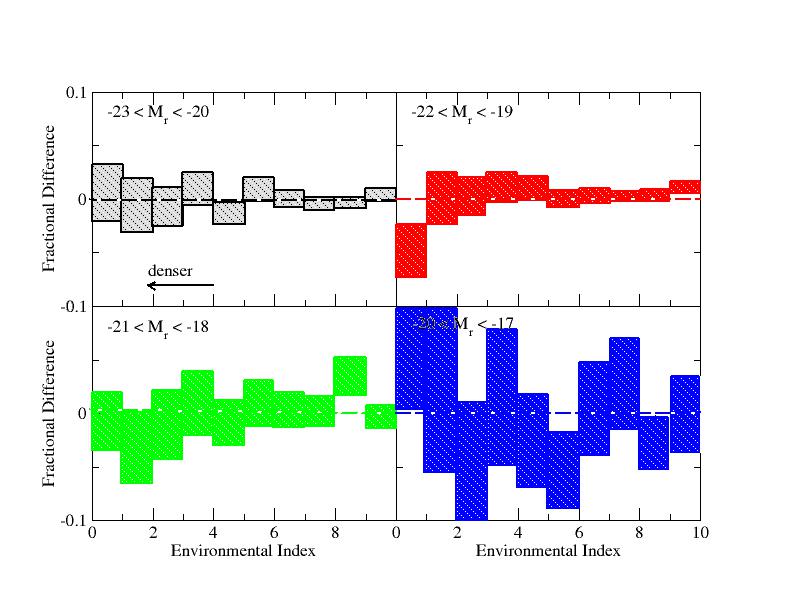}
\caption{Fractional difference in the distribution of environmental densities for galaxies in pairs and the parent sample. A negative
number means that there are fewer pairs, in proportion,  considered, with respect to the reference distribution of
all galaxies in the volume-limited boxes of Fig.~1, in the bin considered, and vice versa. The length of the bar shows the $1\sigma$ range
of the fractional difference while its width shows the range of environmental density in each bin. The arrow in the top left
panel indicates the run of environmental densities sampled by the index.}
\label{fig:fig10}
\end{figure*}

We show the fractional difference of galaxies (in pairs minus all other objects not in close pairs, as with previous figures) in Fig.~\ref{fig:fig10}.  
This does not support the hypothesis that galaxy pairs reside in more dense environments, in agreement with previous work by \cite{Darg2010b} 
and \cite{Ellison2010} that finds only a weak environmental dependence in the SDSS pairs (but see \citealt{Kampczyk2013}, who find a 
much stronger environmental trend, although this is at $z \sim 0.8$ in the COSMOS field). The weak environmental dependence is 
somewhat surprising as simulations indicate a much stronger effect of environmental density than here observed \citep{Jian2012}.

\begin{figure*}
\includegraphics[angle=0,width=\textwidth]{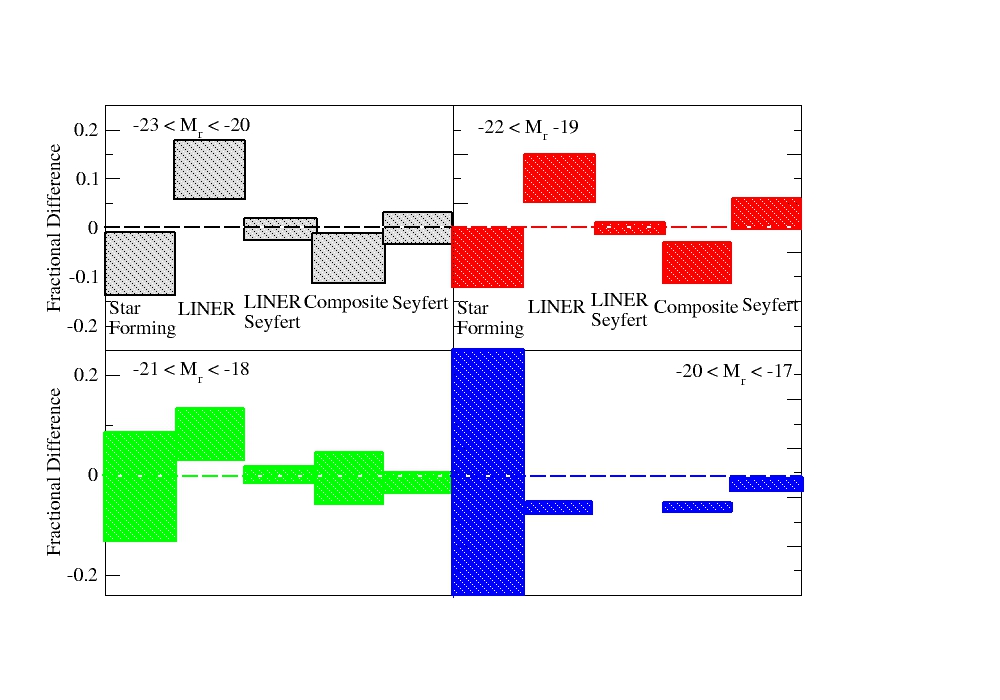}
\caption{Fractional difference in the distribution of BPT classes for galaxies in pairs and the parent sample. A negative
number means that there are fewer pairs, in proportion, in the bin considered, compared to the parent sample of galaxies
in the volume-limited boxes of Fig.~1m and vice versa. The length of the bar shows the $1\sigma$ range
of the fractional difference. The BPT class is indicated in the figure. }
\label{fig:fig11}
\end{figure*}

Finally we consider the BPT diagram (Fig.~\ref{fig:fig11} to measure the degree of activity in galaxies in pairs and their parent
sample. This classifies all emission-line galaxies  as Star-forming, Seyfert, LINER, composite spectra between LINER and Seyfert or 
between Star Forming and Seyfert (here termed Composite) on the basis of emission line intensity ratios. We use the classification of 
\cite{Kauffmann2003} and \cite{Tremonti2004} as applied to SDSS DR8 (and later releases) data by \cite{Thomas2013}. 
This is only a subset of all data in the GAMA survey but it includes about 1/3 of all galaxies in our sample and there is no apparent bias 
towards or against galaxies in pairs. In agreement  with the relative weakness of the  blue peak in the colour distribution of Fig.~\ref{fig:fig8}, 
we find a smaller fraction of  active star-forming galaxies among pairs, but it is surprising to see that there is little or no evidence of increased 
AGN activity among galaxies in pairs compared to the volume-limited parent sample of objects. In all magnitude ranges we consider,
there are fewer Seyfert or Composite (LINER+Seyfert or Star Forming+Seyfert) objects in the pairs sample, while LINERS are only marginally 
more common.

\section{Discussion}

We have measured pair fractions and merger rates for galaxies in the GAMA survey, with $-23 < M_r+5 \log h < -17$ in a series of 
volume-limited samples and explored how the merger rate varies with luminosity, colour or morphology. We then considered how the 
properties of galaxies in close pairs (colour, morphology, environmental density and BPT class) compare to those of more isolated 
objects in the same volume to understand how galaxies are affected by mergers and interactions.

The merger fractions for galaxies in the GAMA survey are (assuming that 50\% of mergers are dynamically bound) $\sim 1.5\%$ almost 
irrespective of luminosity. These are generally in good agreement with previous estimates of the pair fraction and merger rates in the 
local universe, as well as with the well-known flat evolution out to high redshift, as in the compilation by \cite{Lotz2011}.  Our previous
local ($<z> \sim 0.10$) estimate of the merger rate from the MGC \citep{DePropris2005,DePropris2007} is in good agreement with that
presented here, although it concerns galaxies with $-21 < M_B < -18$. \cite{Patton2008} give $N_c=0.021$ but for galaxies with $-22
< M_r < -18$ and only for pairs with luminosity ratios 1:2 or better, compared to our $N_c=0.034$ for galaxies with $-22 < M_r < -19$. 
However, although the pair fraction does not change strongly across this luminosity range, we include more minor mergers are well. 
Approximately 1/2 of our mergers have luminosity ratios of $< 1:2$, which brings the pair fractions into agreement. Comparison of the 
actual volume merger rates requires understanding assumptions concerning the space density of galaxies and assumed merger timescales, 
which are not fully transparent. Finally, the lower redshift points of \cite{Kartaltepe2007} and \cite{Lin2008} are reasonably close to our 
values, despite the somewhat different selection criteria in each case.
 
The volume merger rates are seen to decrease significantly at both high and low luminosities in our sample. Given the measured accretion 
rates and evolution of the merger rate, we estimate that the total stellar mass growth (given the luminosity accretion rates and reasonable 
mass to light ratios) of galaxies in the past 1/2 of the Hubble time due to mergers lies in the range of 10\% to 30\%. This assumes merger 
timescales as discussed above, a 50\% 'true' merger fraction and the flat evolution of the merger rate found in this work to $z \sim 0.2$ 
and measured to $z=1.2$ by other studies (e.g., \citealt{Lin2008, deRavel2009}. \cite{Lopez2012} estimate a similar $\approx 30\%$
for the growth rate of galaxies since $z \sim 1$. However, these tend to be more massive than the GAMA sample which is generally 
between $9 < \log M_*/M_{\odot} < 11$. 

The trend of merger rates with luminosity and redshift is instead not in good agreement with predictions from dark matter simulations 
such as \cite{Murali2002} and \cite{Maller2006}. This is significantly lower than estimates of the mass growth of galactic halos in simulations 
(e.g., \citealt{vandenBosch2002, Cattaneo2011}), which on average double the mass of galaxies since $z \sim 1$. The discrepancy may be 
reduced if the merger rate evolution with redshift is faster but our data for the brighter galaxies are consistent with a flat or declining evolution 
since $z=0.2$, irrespective of colour or morphological type, as previously observed for similarly luminous galaxies at higher redshifts by 
\cite{Bundy2004,Lin2004,Lin2008} and \cite{deRavel2009}. In the compilation of \cite{Lotz2011}, the merger rate evolves as $(1+z)^{\sim 0.1}$ 
for luminosity (or mass) selected samples, in agreement with very slow growth rates by major merging for massive galaxies. Robotham et al. 
(2014, MNRAS submitted) also find that the evolution of the merger rate is close to flat. 

Our data allow us to consider the local dry, wet and damp merger rate. Dry mergers (between either red or spheroid-dominated galaxies) are 
observed to be rare in the local universe, in agreement with several previous studies \citep{Masjedi2006,Masjedi2008,Wen2009,
Robaina2010,Jiang2012}, while we also observe that dry mergers are less frequent as a function of decreasing luminosity. We find evidence 
of slow evolution for at least the more massive (luminous) systems. In our close pair study of luminous red galaxies at $<z> \sim 0.55$ 
we measured a stringent upper limit of $< 0.8\%$ per Gyr, which is consistent with flat evolution of the dry merger rate out to at least $z=0.6$. 
\cite{Chou2012} also confirm the slow growth of the dry merger rate as a function of redshift out to $z=0.7$.  It is therefore unlikely that the 
red sequence is formed by dry mergers at least in the past 1/3 of the Hubble time. Most mergers between lower luminosity galaxies are instead 
wet (and the rate increases towards lower luminosities) while damp mergers are intermediate.

When we compare the colour distribution of galaxies in pairs with that of galaxies in their volume-limited parent sample, we observe a deficit
of blue objects and an excess of red galaxies. This argues for suppression of star formation in galaxy pairs, as noted earlier by \cite{Darg2010a}
and \cite{Ellison2010}. This is somewhat surprising in the light of theoretical expectations that mergers and interactions boost the star formation
and activity rates (e.g., \citealt{Mihos1996}). We confirm this from our analysis of BPT classes where there is a lower fraction of star-forming 
galaxies among close pairs compared to galaxies in the same volume. \cite{Bergvall2003} found only a modest degree of enhancement of 
star formation for interacting pairs, contributing little to the luminosity; \cite{Lin2007} also find only a modest enhancement of star formation in 
galaxy pairs in a wide range of merger stages; \cite{Li2008} instead find that the enhancement in star formation, in a sample of star-forming
galaxies, increases with decreasing separation but does not depend on the luminosity ratio, although it is stronger for lower-mass galaxies;
for massive galaxies, \cite{Robaina2009} find that only 10\% of star formation is due to major interactions (albeit at higher redshifts than 
studied here); \cite{Patton2011} also detect an excess of red galaxies in pairs and attribute this to denser environments and hence more
significant spheroids (because of the morphology-density relation); \cite{Chou2012} confirm that most close pairs are red and not blue; 
\cite{Deng2013} also finds evidence that interactions are not generally responsible for star formation bursts in two volume-limited samples 
drawn from the SDSS; \cite{Cluver2014} study star formation for galaxies in groups and claim that galaxies with a close neighbour have lower 
star formation rates. Similarly, we find that, by comparing the distributions of S{\'ersic} indices for galaxies in pairs and their parent samples, there 
is a relative lack of 'disks' and excess of 'spheroids' among paired galaxies as in \cite{Patton2011}. spheroids are generally associated with suppression 
of star formation, but it is not clear whether this a cause or an effect \citep{McGee2011}. If pairs lie in more dense environments, they will 
contain more prominent spheroids (by the standard morphology-density relation) and therefore have lower star formation rates (as suggested 
by \citealt{Patton2011}, explaining the excess of red galaxies among pairs found above. However, in our sample, we do not find that galaxies in 
pairs lie in significantly different environments (as do \citealt{Darg2010b} and \citealt{Ellison2010} who find only a mild environmental dependence). 
This suggests a degree of morphological evolution is also associated with star formation suppression in these objects. \cite{Li2008} show 
that tidal interactions may indeed cause an increase in galaxy concentration, which is an indicator of more significant spheroids. However, this is 
not due to the fact that the galaxies lie in more dense environments and are more spheroid-dominated (and therefore more quiescent). Rather, 
it appears that the extra spheroid light is related to the suppression of star formation and presence in a close pair, although it is not possible to 
determine from the present sample whether quenching or morphological evolution occur first \citep{Li2008}.

One mechanism by which we may explain these findings is `harassment' \citep{Moore1996}. Galaxies in these close pairs are likely to have 
undergone a series of previous encounters and close passes; \cite{Patton2013} argues that about 2/3 of the star formation in close pairs takes
place at separations above 30 kpc. This leads to a gradual suppression of star formation as gas is exhausted or driven out, while dynamical 
relaxation leads to secular evolution of spheroids without mergers. \cite{Chou2012} also suggest that in their sample star formation is triggered at 
large radii and then suppressed to form red close pairs. \cite{Robotham2013} point out the role of mergers and interactions in 
suppressing star  formation in most galaxy pairs, while \cite{Trinh2013} present an analysis of the properties of single and widely spaced 
paired galaxies in the SDSS and find evidence of delayed quenching, over timescales of a few Gyrs, consistent with a `harassment' scenario 
as proposed here. On the other hand, an Integral Field Unit study of the Mice pair, shows that little star formation has been induced so
far \citep{Wild2014}.

The BPT classifications, from emission line diagnostics, support the observation that galaxies in pairs are less active than their counterparts
in the same volume. As remarked above, there is a deficit of star-forming galaxies in pairs. Surprisingly, there is also no evidence of excess 
AGN activity in galaxies in pairs. This is surprising as one expects mergers to trigger nuclear activity \citep{Mihos1996,Springel2005,
Hopkins2008}. In these models AGNs occur at or shortly after final coalescence so we should expect double QSOs in close pairs 
\citep{Comerford2009}. \cite{Bessiere2012}, however, find QSOs at several stages of the merging process even before the final merger. 
On the other hand, \cite{Cisternas2011} show that AGNs do not lie in more disturbed hosts and conclude that mergers are not necessarily
responsible for nuclear activity. \cite{Teng2012} search for double QSOs in pairs of galaxies where one member is already known to host 
an AGN and find only one such case out of twelve objects. This suggests that nuclear activity is not necessarily related to the on-going 
mergers or interactions in these objects. \cite{Ellison2008} also reports on a relative lack of AGN activity in close pairs from the SDSS
although \cite{Patton2011} claims to find some enhancement from the same sample. \cite{Deng2013b} also argue that there is less 
than 1$\sigma$ significance for AGN triggering in mergers from a volume-limited sample of SDSS galaxies. We find no evidence of 
AGN activity in strongly interacting mergers and merger remnant post-starburst galaxies in \cite{DePropris2014}, although there are indications of centrally 
concentrated star formation and inside-out quenching in these objects. Other studies find a stronger correlation between presence in a close 
pair and AGN activity \citep{Patton2011}. In some cases, these discrepancies may be due to issues of selection, including the use of 
emission-line diagnostics rather than X-ray or mid-IR fluxes. 

At face value our findings minimise the importance of major mergers and interactions for galaxy evolution and argue that most 
galaxy evolution takes place via internal and secular processes, as well as minor mergers, at least at low redshift. This is also the 
conclusion of several studies in this field (e.g., \citealt{Lopez2012,Huang2013,Kaviraj2013,McLure2013,Fritz2014}) and is broadly 
in agreement with more recentmodels where minor mergers and secular evolution may play more important roles in galaxy evolution
in the past 1/2 of the Hubble time (e.g., \citealt{Guo2008,Parry2009}).


\end{document}